\documentclass{PoS}

\title{
\begin{picture}(0,0)(0,0)%
   \put(230,75){\makebox(0,0)[l]{\textnormal
{\normalsize OU-HET-832, KEK-CP-313, YITP-14-85%KEK-CP-???
}}}%
\end{picture}%
Topology density correlator on dynamical domain-wall ensembles with nearly frozen topological charge}

\ShortTitle{Topology density correlator on dynamical domain-wall ensembles}

\author{JLQCD Collaboration: 
        \speaker{H.~Fukaya}$^a$\thanks{E-mail: hfukaya@het.phys.sci.osaka-u.ac.jp},
        S.~Aoki$^{b}$,
        G.~Cossu$^{c}$,
        S.~Hashimoto$^{c,d}$,
        T.~Kaneko$^{c,d}$,
        J.~Noaki$^{c}$,
        \\
        \\
        \\
        \llap{$^a$}
        Department of Physics, Osaka University, 
        Toyonaka, Osaka 560-0043 Japan
        \\
        \llap{$^b$}
        Yukawa Institute for Theoretical Physics, Kyoto University, Kyoto 606-8502, Japan
        \\
        \llap{$^c$}
        High Energy Accelerator Research Organization (KEK),
        Tsukuba 305-0801, Japan 
        \\
        \llap{$^d$}
        School of High Energy Accelerator Science,
        The Graduate University for Advanced Studies
        (Sokendai),Tsukuba 305-0801, Japan
}

\abstract{
  Global topological charge decorrelates very slowly or even
freezes in fine lattice simulations. On the other hand, its local 
fluctuations are expected to survive and lead to the correct physical
results as long as the volume is large enough. 
We investigate this issue on recently generated configurations including dynamical
domain-wall fermions at lattice spacings $a=0.08$ fm and finer. We utilize the Yang-Mills
gradient flow to define the topological charge density operator and
calculate its long-distance correlation,
through which we propose a new method for extracting 
the topological susceptibility in a sub-volume.
This method takes care of the finite volume correction, 
which reduces the bias caused by the global topological charge.
Our lattice data clearly show a shorter auto-correlation time 
than that of the naive definition using the whole lattice,
and are less sensitive to the global topological history.
Numerical results show a clear sea-quark mass dependence,
which agrees well with the prediction of chiral perturbation theory.
}

\FullConference{The 32nd International Symposium on Lattice Field Theory,\\
		23-28 June, 2014\\
		Columbia University New York, NY}

\begin{document}

\section{Introduction}

The configuration space
of continuum QCD  is not smoothly connected, 
and divided into  sectors characterized by the 
global topological charge $Q$. 
They are separated by infinitely high barriers
of the gauge action.
On the lattice, the barriers become finite, and the Monte Carlo simulation
can sample the configurations at different topological charges.
As approaching the continuum limit, however,
the barriers grow rapidly so that the hybrid Monte Carlo updates 
cannot frequently go across these topological boundaries.
As a consequence, the auto-correlation time of topological charge
becomes very long already at a lattice spacing $a\sim 0.05$ fm
\cite{Schaefer:2010hu, Bruno:2014ova}.

In our previous simulations \cite{Aoki:2007pw,Chiu:2008kt,Cossu:2013uua} 
with the overlap quarks,
we took this property as an advantage 
and performed the QCD simulations in fixed topological sectors.
Avoiding the topological boundaries is effective to reduce the
numerical cost for the overlap quark determinant
which has discontinuities on the boundaries.
This can be achieved by introducing an extra Wilson fermion determinant
with a negative cut-off scale mass, which prevents
the index of the overlap operator to change along the Monte Carlo history \cite{Fukaya:2006vs}.
We developed a method to correct the
fixed topology effects, and succeeded in
extracting the topological susceptibility 
from the local topological fluctuation 
in the simulation with a fixed global topological charge \cite{Aoki:2007pw,Chiu:2008kt}.

We have recently launched a new project of simulating QCD
with chiral fermions on finer and larger lattices \cite{Kaneko:2013jla}. % Ref Kaneko
Our goal is to cover the lattice size of around 
2.5--4 fm with the cut-off $1/a\sim$ 2.4--4.5 GeV.
We employ the M\"obius domain-wall fermions \cite{Brower:2004xi}
% M\"obius
which is numerically less expensive than the overlap fermions,
%Although the M\"obius domain-wall Dirac operator has 
%no {\it exact} chiral symmetry, 
while keeping %It still keeps 
the violation of the
Ginsparg Wilson relation still small at the $0.1$\% level
compared to the cut-off.
It turned out that this size of violation is sufficient
to smooth out the quark determinant and the hybrid Monte Carlo
has no obstacle in crossing the barriers.
%We have thus obtained a way of simulating QCD with good chirality,
%keeping the topology tunnelings active.
%However, as the lattice spacing decreases, we observe again
%a long auto-correlation of the global topological charge, as is expected.
The problem of the long autocorrelation time exists as with 
other fermion formulations since the topology barriers due to the gauge action remains.

Since %we have learned from our fixed topology study 
%\cite{Aoki:2007pw,Chiu:2008kt} that
the effect of the global topological charge is 
nothing but a finite volume effect \cite{Brower:2003yx, Aoki:2007ka}, 
once it is removed, the auto-correlation associated with it could also be removed.
%We have also learned that it is more important 
%to investigate the local or sub-volume
Since the physical effect of the topological charge should be found 
in its local excitations
\cite{deForcrand:1998ng, Brower:2014bqa},
in this work, we consider the local fluctuation of topology
using the topological charge density operator constructed via the gauge links after performing
the Wilson flow.
As shown in Ref.~\cite{Luscher:2010iy}, gluonic quantities after the gradient flow
are free from UV divergences, and the gluonic 
definition of the topological charge density 
is closer to the continuum limit as the gauge fields become
smoother after the flow. 

With this gluonic construction, we propose a new
method for extracting the topological susceptibility.
Its definition is given in a sub-volume, and contains an $1/V$ correction term,
to reduce the bias caused by the global topological charge.
Our lattice data show a much shorter auto-correlation time
than that of the naive calculation of the topological susceptibility 
summed over the whole lattice sites.
We also find that our data indeed cancel the bias from the global topological charge.
Moreover, the numerical results show a clear sea-quark mass dependence,
which agrees well with the prediction of chiral perturbation theory.

\section{A new method for extracting the topological susceptibility}

%It can be generally shown 
In a finite volume $V$ 
the effect of the global topological charge
$Q$ to any quantity is given by a series of $Q^2/V$, as well as $1/\chi_t V$ 
provided that the volume is sufficiently larger
than the inverse of the topological susceptibility $\chi_t$.
More explicitly,  the expectation value of any CP-even operator
$O$ in a fixed topological sector of $Q$
can be expressed by a series \cite{Brower:2003yx, Aoki:2007ka}
%%% Eq O_Q general form
\begin{eqnarray}
\langle O \rangle_Q = \langle O \rangle_{\theta=0} + 
\left.\frac{\partial^2}{\partial \theta^2}
\langle O \rangle_\theta\right|_{\theta=0}\times \frac{1}{2 \chi_t V}
\left[1-\frac{Q^2}{\chi_t V}\right]+\cdots,
\end{eqnarray}
when $\chi_t V\gg 1$.
This is intuitively understood from the clustering property
of quantum field theory that only nearby 
region can affect the local observables.
%Thus, it is enough for any physical observables to
Therefore, it is sufficient to measure any physical observable in a sub-volume,
as long as the size of the sub-volume is larger
than the correlation length of the system.
%, {\it e.g.} $1/m_\pi$ in QCD.
Even the topological susceptibility is not an exception.

Suppose we have a good definition of the 
topological charge density operator $q(x)$.
The topological susceptibility is conventionally defined by
%%% Eq chi_t naive 
\begin{equation}
\chi_t = \int d^4 x \langle q(x)q(0)\rangle.
\end{equation}
Using the clustering property and the fact that the lowest energy state
couples to $q(x)$ is the eta-prime meson 
(with the mass $m_{\eta^\prime}$), 
one can truncate the integral at some radius $r_{\rm cut}(>1/m_{\eta^\prime})$:
\begin{equation}
\int_{|x|<r_{\rm cut}} d^4 x \langle q(x)q(0)\rangle 
= \chi_t + \mathcal{O}(e^{-m_{\eta^\prime}r_{\rm cut}}).
\end{equation}

When the configurations are generated in a fixed topological sector
or in various sectors but their distribution is rather biased,
there are potential effects from the badly sampled $Q$ as $1/V$ corrections.
Thus, it is better if one can subtract this correction in advance.
Luckily, in the case of topological susceptibility, one can 
uniquely determine this correction term with no free parameter to tune.
Namely, at long distances $|x|>r_{\rm cut}$, 
the correlator $q(x)q(0)$ in the topological sector of $Q$
is determined purely by the finite volume effects due to $Q$ \cite{Aoki:2007ka},
%%% Eq qq correlator %%%%%%%%
\begin{equation}
\label{eq:qqlong}
\langle q(x)q(0) \rangle_Q \sim \frac{1}{V}\left[\frac{Q^2}{V}-\chi_t\right].
\end{equation}
Even when the sampling of the configuration has a bias 
in the global topological charge $Q$, its effect can be subtracted,
according to this formula. This is achieved by calculating
%Now we give a new definition of the topological susceptibility\footnote{
%Here, we put the source at the origin 
%but numerically we average over different source points.}
%by
%%% Eq chi_t bar %%%%%%%%%%%%%
\begin{equation}
\label{eq:chitbar}
\bar{\chi}_t = \frac{V}{V-V_{\rm sub}}\left\langle 
\int_{|x|<r_{\rm cut}} d^4 x\;q(x)q(0) -\frac{V_{\rm sub}}{V^2}Q^2\right\rangle, 
\end{equation} 
where $V_{\rm sub}$ is the volume of the sub-domain in the range 
$|x|<r_{\rm cut}$. 
Note that $\chi_t = \bar{\chi_t}$ when the sampling of the topology has no bias.
%We expect a lot of advantages of $\bar{\chi_t}$ compared to 
%the naive definition $\chi_t$.
Since $\bar{\chi}_t$ is defined in a sub-volume, 
we can use the translational invariance and average 
it over the whole volume, which reduces the statistical uncertainty,
and assures that the smooth limit $\displaystyle \lim_{V_{\rm sub}\to V}\bar{\chi_t}=\chi_t $ 
even for finite statistics of the gauge samples.

We expect that $\bar{\chi}_t$ has a shorter auto-correlation time
than $\chi_t$ for two reasons. First, the topological lump can freely 
go in and out of the sub-domain whose moving time-scale
is reported to be much shorter than $Q$ \cite{McGlynn:2014bxa}.
Second, the bias from global topology history is removed in advance,
and the measurement should be insensitive to the (biased) history of
the global topology.

The above observation is expected to be true only when 
we have a good definition of $q(x)$.
For example, a naive pseudoscalar operator 
made from the plaquette on the generated configuration
does not work very well. This is a part of the reason why 
conventionally the topological susceptibility has been measured by
the global topological charge as $\langle Q^2\rangle/V$.
In this work, we apply the Wilson flow cooling
and use a naive gluonic definition of the topological charge density
after the flow \cite{Luscher:2010iy}.
Any gluonic operator on the configurations cooled by the gradient flow 
has been shown to be free from UV divergences \cite{Luscher:2010iy}.
Moreover, if the plaquette values become sufficiently smooth,
one can uniquely determine the global topological charge
by a gluonic quantity.
As we will see below, $q(x)$ after the gradient flow
at the smearing range $\sim 0.5$ fm shows the expected properties.

\section{Lattice simulation and Yang-Mills gradient flow}

For the configuration generation,
we employ the Symanzik gauge action and 
the M\"obius domain-wall fermion action.
The determinant of the M\"obius domain-wall Dirac operator
is equivalent to 
that of an approximation of the overlap Dirac operator:
%%% Eq 4d DDW %%%%%%%%%%%%%%%
\begin{eqnarray}
D(m) &=& \frac{1+m}{2}+\frac{1-m}{2}\tanh(L_s \tanh^{-1}(2H_T)),\;\;\;\;
%\naonumber\\
2H_T = \gamma_5\frac{2D_W}{2+D_W},
\end{eqnarray}
where $L_s$ is the size of 5-th direction,
$D_W$ is the Wilson Dirac operator,
and $m$ is the quark mass.
We apply three steps of stout smearing 
of the gauge link before inserting it
in the Dirac operator.

We carry out $2+1$-flavor lattice QCD simulations on two 
different lattice volumes
$L^3\times T(\times L_s)$=$32^3\times 64(\times 12)$
and $48^3\times 96(\times 8)$, for which
we set $\beta=4.17$ and 4.35, respectively.
The lattice spacings are estimated to be 
2.4 GeV and 3.6 GeV, respectively, and therefore, 
the two lattices share a similar physical size.
For the quark mass, we use two values of 
the strange quark mass $m_s$ around its physical point, 
and 3--4 values of the up and down quark mass $m_{ud}$ for each $m_s$.
The lightest pion mass is around 220 MeV with 
our smallest value of $m_{ud}=0.0035$.

\begin{figure*}[tb]
  \centering
  \includegraphics[width=7.5cm]{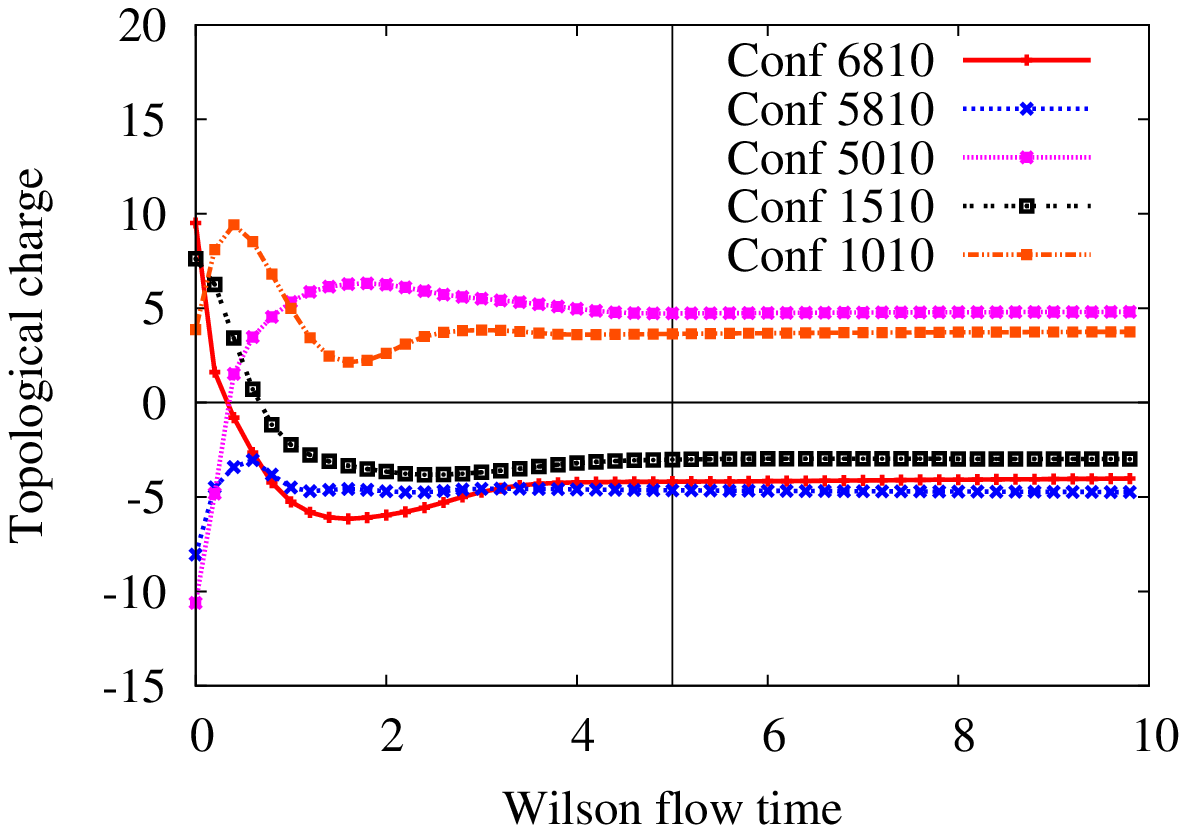}
  \includegraphics[width=7.5cm]{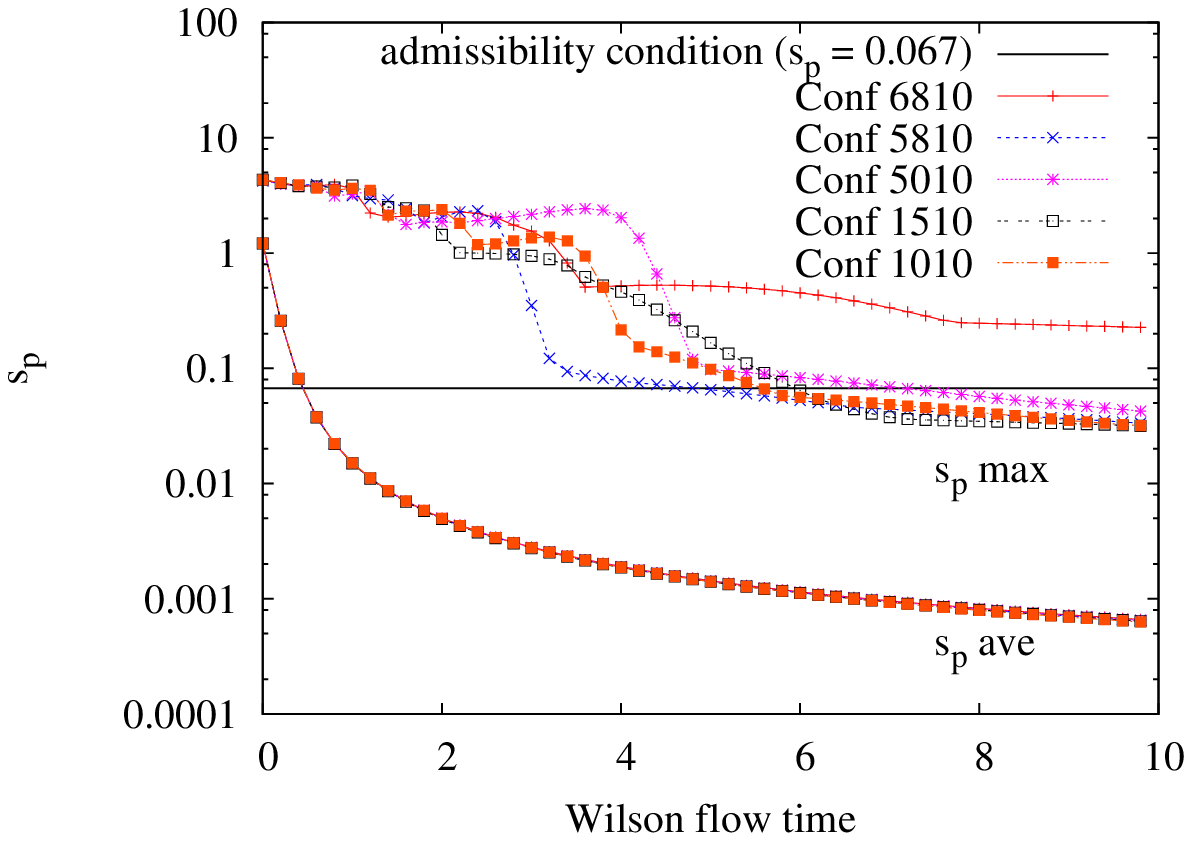}
  \caption{Wilson flow history of the topological charge $Q$ (left panel), 
   and the average and maximum of $s_p$ (right) for typical 5 configurations
   generated at $\beta=4.17$, $m_{\rm ud}=0.007$, and $m_s=0.030$.
   At the flow time of $t=5$ ($\sim 1.4 t_0$), the topological charge 
is well saturated.
    }
  \label{fig:Wflow}
\end{figure*}

For each configuration to be measured, 
we perform 500--1000 steps of the Wilson flow
with a step-size 0.01. 
Fig.~\ref{fig:Wflow} shows the Wilson flow time history
of the gluonic definition of the topological charge $Q$ (left panel),
and the average and maximum of $s_p={\rm Re\;Tr}[1-U_p]$, 
where $U_p$ denotes the plaquette (right panel).
These plots represent typical 5 configurations
generated at $\beta=4.17$, $m_{\rm ud}=0.007$, and $m_s=0.030$.
It is known that if every plaquette satisfies the condition $s_p < 0.067$,
there exists a well-defined topological charge \cite{Luscher:2010iy}.
As seen in the figure, after the flow time of $t=5$,
the topological charge does not change,
and the average of $s_p$ is well below the condition $0.067$
(although a few plaquettes are still above this value). 
%We also find a good agreement ($\sim$80--90\%)of $Q$ with 
%the index of the overlap Dirac operator on 
%smaller $L^3\times T=16^3\times 8$ lattices, 
%which is performed for the finite temperature study \cite{Tomiya}. % Ref Tomiya

The flow time $t=5$ (for $\beta=4.17$) corresponds to $1.4 t_0$, where
$t_0$ is the reference scale $\sim 0.0236$ fm$^2$.
The smeared region is then around $\sqrt{8t}\sim 0.5$ fm. 
We should, therefore, be able to extract
the local topological fluctuation 
when the sub-domain size is larger than 0.5 fm.
In the following analysis, we compute the
topological density correlators at $t_{\rm ref}=5$ (for $\beta=4.17$)
and $t_{\rm ref}=10.8$ (for $\beta=4.35$ runs) whose 
physical sizes are roughly equal.

\section{Preliminary results}

\begin{figure*}[tb]
  \centering
  \includegraphics[width=7.5cm]{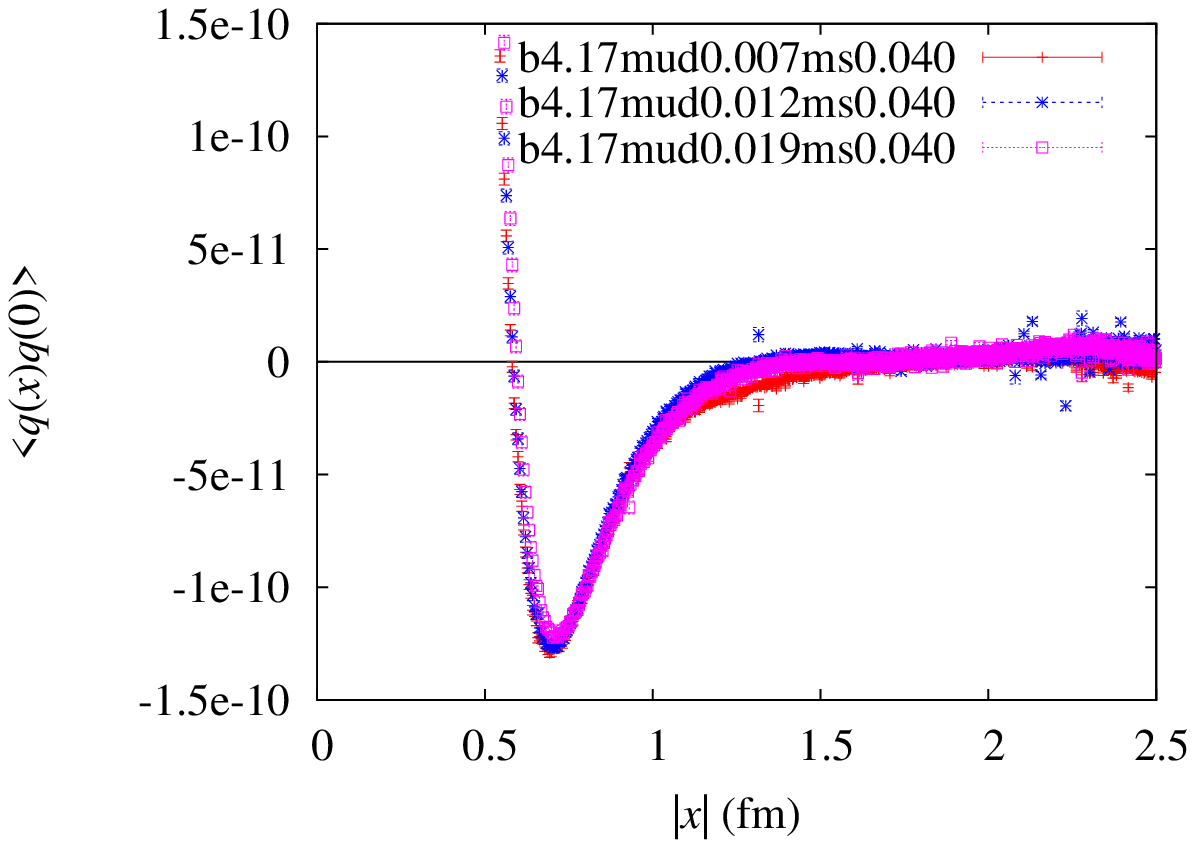}
  \includegraphics[width=7.5cm]{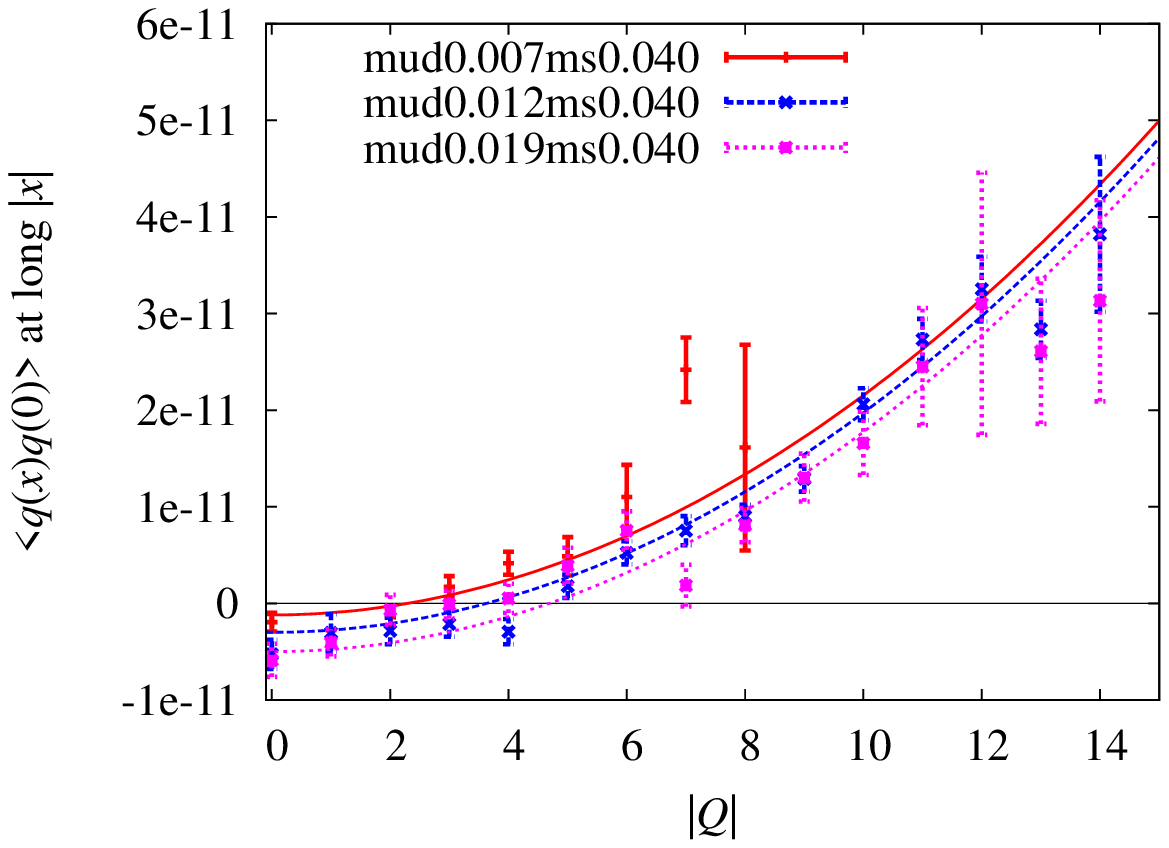}
\caption{
    Topological charge density correlator (left panel) 
    and its $|Q|$-dependence (right) 
    in the long distance region  $r_{\rm cut}< |x| < L$ at $\beta=4.17$.
    }
  \label{fig:topcorr}
\end{figure*}

First, we compute the topological charge density correlator $\langle q(x)q(0)\rangle$
at the Wilson flow time $t_{\rm ref}$ as a function of $|x|=\sqrt{x^2}$.
Different points giving the same $|x|$ are averaged.
The left panel of Fig.~\ref{fig:topcorr} shows the data from
the $\beta=4.17$ runs.
The correlator has a positive core 
in the short distance region, and 
goes to negative in the intermediate region,
and comes back to around zero.
%which behavior is well-known in the literature \cite{Bruno:2014ova}.
We find saturation (to zero) around $|x|=$1.5--2 fm.
We therefore set $r_{\rm cut}=1.6$ fm
as the truncation length for %our new definition of 
the topological susceptibility calculated using Eq.~(\ref{eq:chitbar}).

By also measuring $Q=\sum_x q(x)$, we divide the data
into topological sectors and check their $Q^2$ dependence 
at long distances, which is predicted in Eq.~(\ref{eq:qqlong}). 
The right panel of Fig.~\ref{fig:topcorr} shows
its average in the range $r_{\rm cut}< |x| < L$.
Note that the data at $x_\mu > L/2$ for any $\mu$ are not averaged
to avoid possible effects of the boundary.
The expected dependence $Q^2/V^2$ is clearly seen. %, as predicted.
We emphasize that this quadratic
function has no free parameter to tune.
In this plot, we also draw curves whose intercept
is given by $\bar{\chi_t}$ determined below.

\begin{figure*}[tbh]
  \centering
  \includegraphics[width=7.5cm]{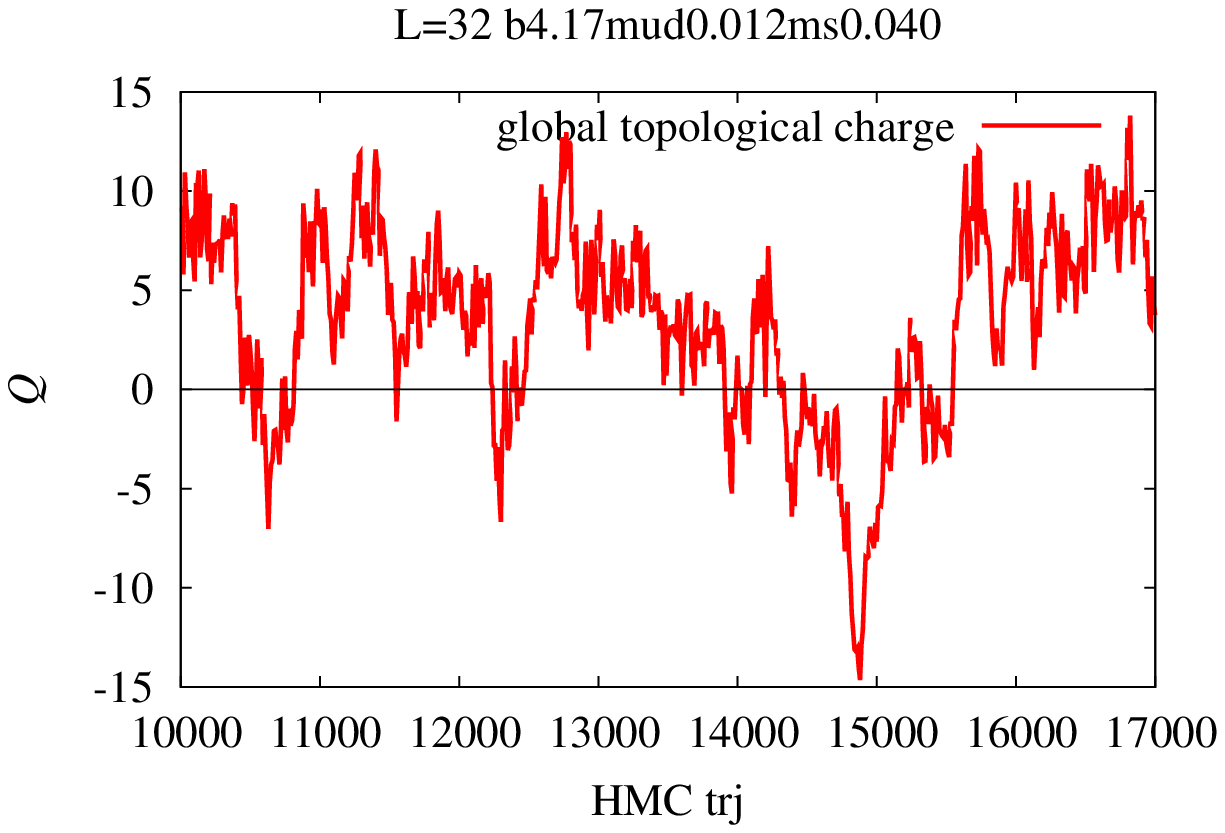}
  \includegraphics[width=7.5cm]{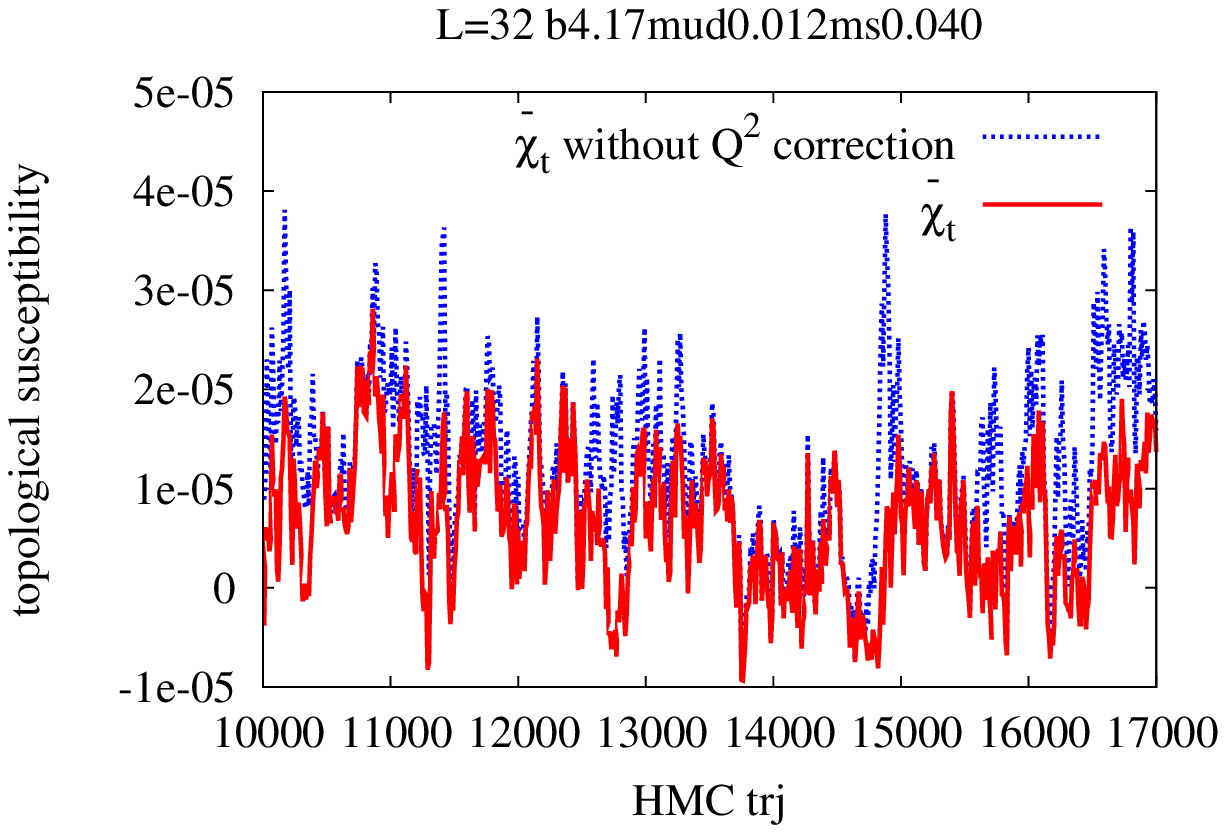}  
  \includegraphics[width=7.5cm]{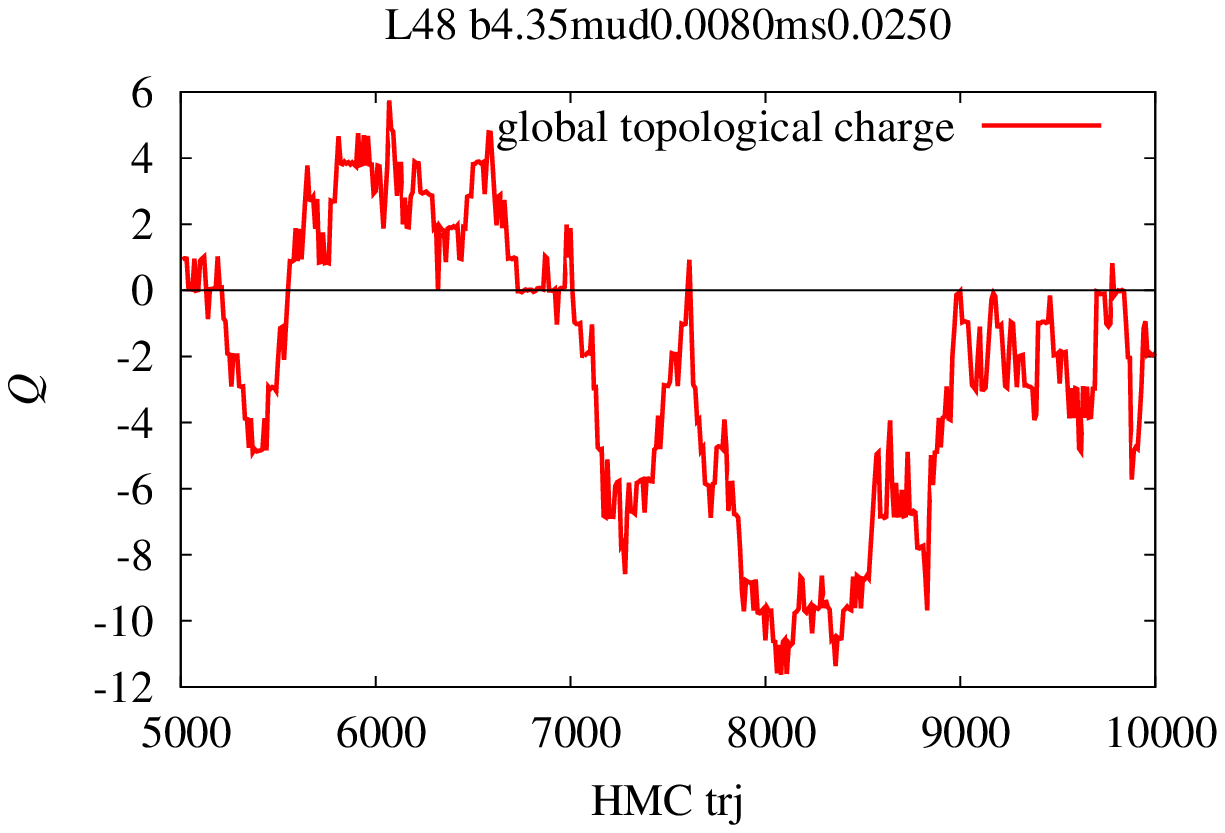}
  \includegraphics[width=7.5cm]{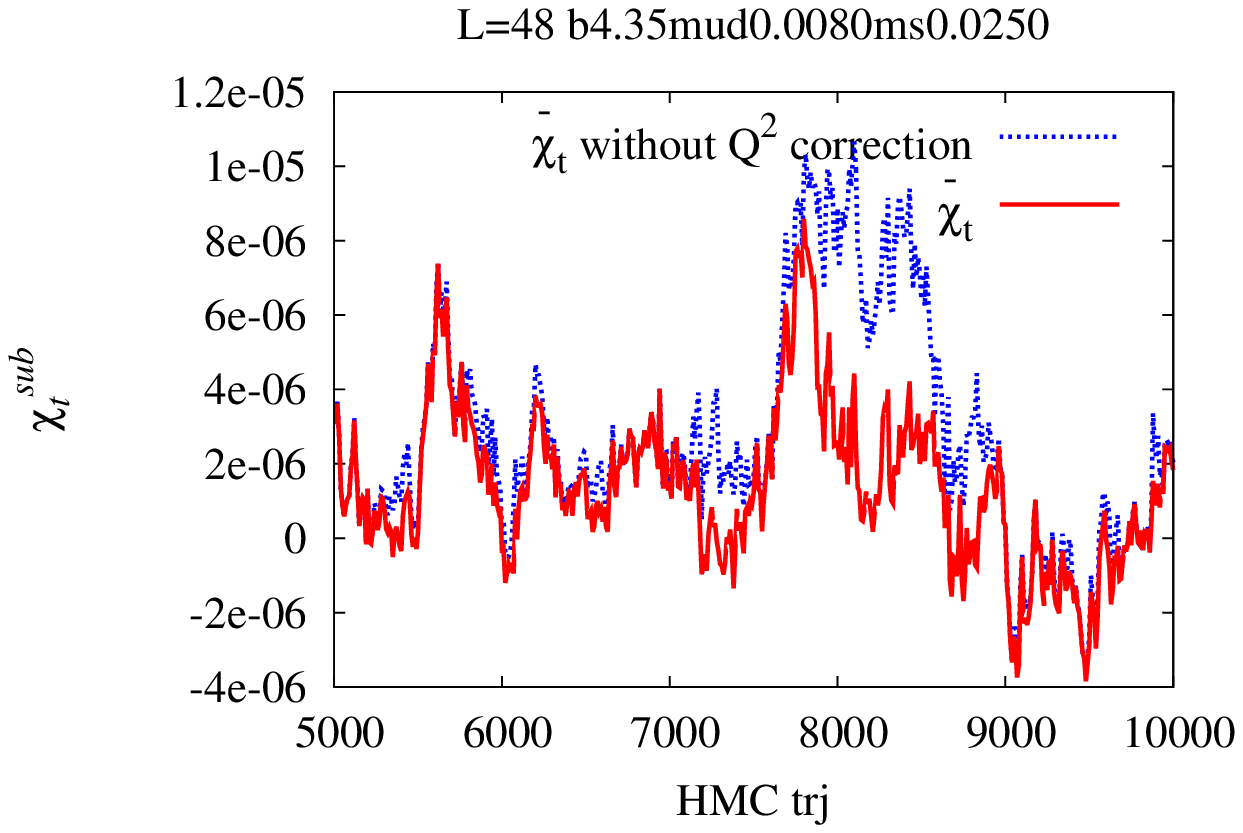}  
\caption{
  Monte Carlo history (solid lines) of the global topological charge (left)
  and our new definition of the topological susceptibility $\bar{\chi}_t$ (right)
  at $\beta=4.17, m_{ud}=0.012, m_s=0.040$ (top) and $\beta=4.35, m_{ud}=0.0080, m_s=0.0250$ (bottom).  
  We also plot  $\bar{\chi}_t$ without the $Q^2/V$ correction term 
(dotted lines).
%  for which the correlation with the bias in the global topology looks stronger. 
    }
  \label{fig:qhistory}
%\end{figure*}
%\begin{figure*}[tbh]
  \centering
  \includegraphics[width=7.5cm]{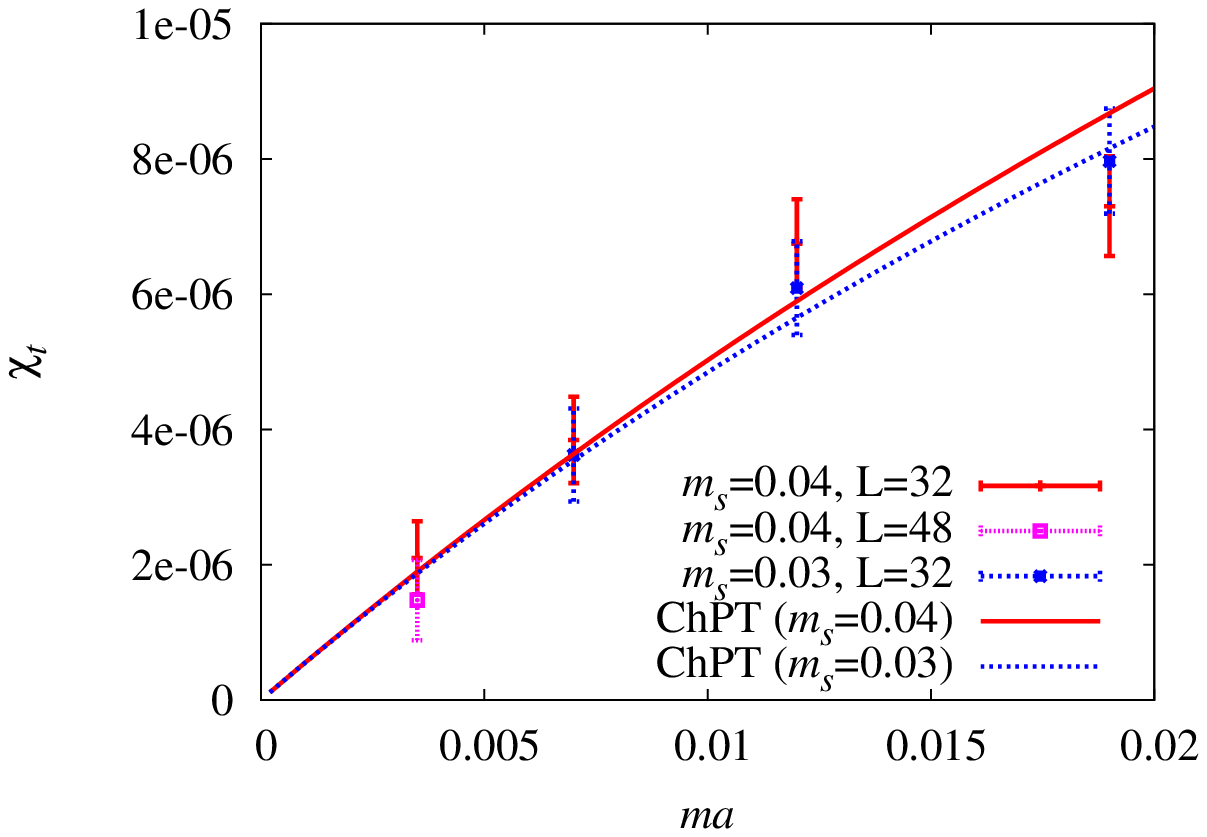}
  \includegraphics[width=7.5cm]{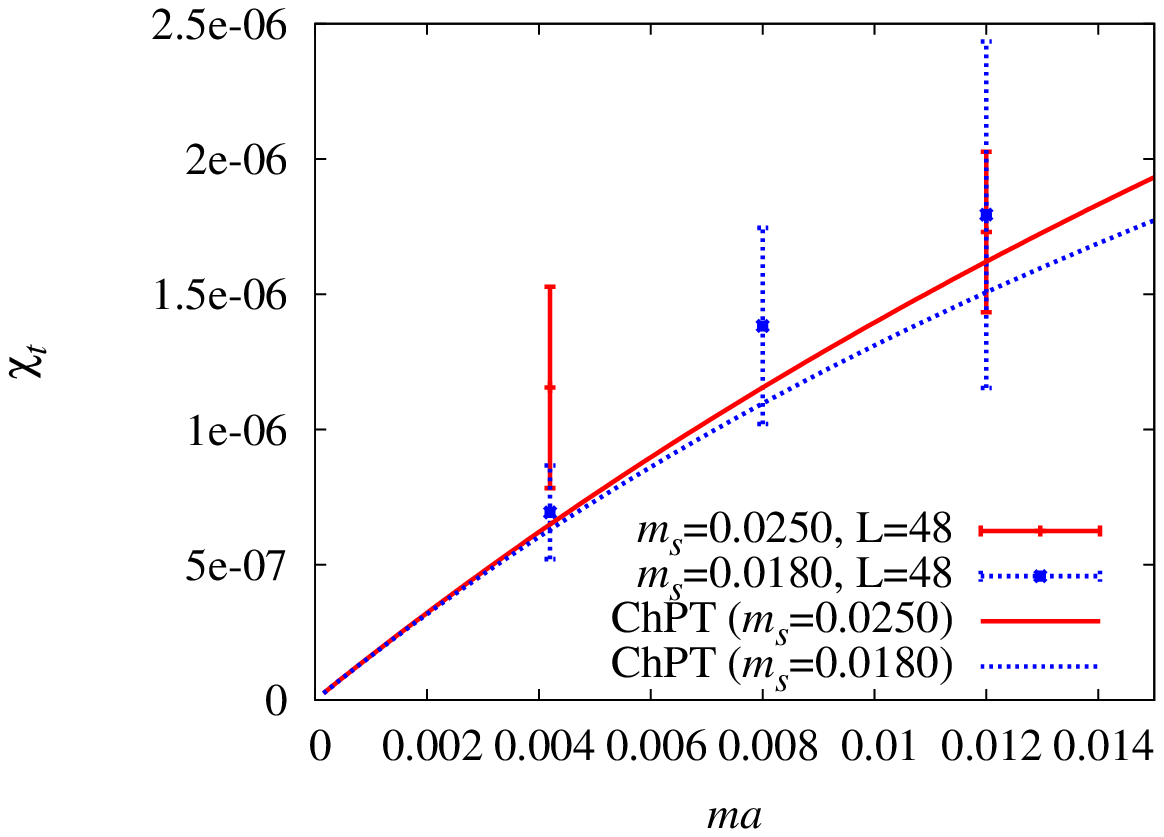}
\caption{
  The $m_{ud}$ dependence of $\bar{\chi}_t$ at $\beta=4.17$ (left)
  and $\beta=4.35$. For a comparison, we draw the ChPT prediction $\chi_t^{\rm ChPT}$
  with an input of the chiral condensate $\Sigma=(250\;\mbox{MeV})^3$.
    }
  \label{fig:topsus}
\end{figure*}

We are now ready for computing $\bar{\chi_t}$ in Eq.~(\ref{eq:chitbar}).
%with $r_{\rm cut}=$ 1.6 fm.
Note that for our  choice $r_{\rm cut}=$ 1.6 fm, $V_{sub}/ V\sim $30 \%.
Namely, the instanton-like lump has enough space, 
70\% of the whole volume, to escape.
Here again the data at $x_\mu > L/2$ for any $\mu$ are not averaged
to avoid the effect of the boundary.
As Fig.~\ref{fig:qhistory} shows, the Monte Carlo history 
of %the new definition of the topological susceptibility 
$\bar{\chi_t}$ 
(right panels) fluctuates more frequently than the 
global topological charge (left panels).
In the same panel, we also plot $\bar{\chi_t}$ without 
the correction term from the global topological charge,
which apparently shows a stronger correlation with 
the global topological charge.
Namely, this term plays the expected role of canceling 
the bias from the global topology.

Moreover, as shown in Fig.~\ref{fig:topsus}, 
we find that the sea quark mass dependence of 
$\bar{\chi_t}$ agrees well with the prediction from 
chiral perturbation theory,
%%% Eq chi_t ChPT
\begin{eqnarray}
\label{eq:chitCHPT}
\chi_t^{\rm ChPT} = \frac{\Sigma}{1/m_u + 1/m_d + 1/m_s},
\end{eqnarray}
where $\Sigma$ denotes the chiral condensate.
We can even estimate the (bare) value of chiral condensate 
as $\Sigma=(250\;\mbox{MeV})^3$.
%Such a remarkable sea quark dependence 
%is not found in the literature \cite{Bruno:2014ova}. %% Ref.
Our definition $\bar{\chi_t}$ on the configurations generated by
the  M\"obius domain-wall fermions seems to 
correctly reflect the sea quark's effect. % from the gluonic quantity.
This implies that the topological fluctuation in our ensembles
is just as expected from the effective theory approach.
It should be noted that $\bar{\chi}_t$ is constructed
purely from gluonic quantities. The fermion loop effect is
clearly visible in the gauge sector,
thanks to the clean signal after the Wilson flow and to the
good chiral symmetry of the M\"obius domain-wall fermion.\\

Numerical simulations are performed on IBM System Blue Gene Solution at KEK 
under a support of its Large Scale Simulation Program 
(Nos. 12/13-04 and 13/14-04). 
%We thanks H. Matsufuru for the support on the computing facility and P. Boyle 
%for helping in the optimization of the code for BGQ. 
This work is supported in part by the Grand-in-Aid of 
the Japanese Ministry of Education (No.25800147, 26400259 26247043),
%the Grant-in-Aid for Scientific Research on Innovative Areas (No.),
the Grant-in-Aid for Scientific Research (B) (No. 25287046), 
and SPIRE (Strategic Program for Innovative Research) Field 5.


%\section{Summary}

%We have studied the local fluctuation of topology 
%on the gauge ensembles generated by the dynamical domain-wall quarks.
%On the configurations cooled down with the Wilson flow,
%where the smeared size is expected to be $\sqrt{8t}\sim 0.5$ fm,
%the naive gluonic definition of the topological charge density 
%operator shows an expected behavior : having a well determined 
%global topological charge, two-point correlation saturated around 1.6 fm,
%and even the $Q^2/V$ effect from the global topological charge.

%With these observations, we have proposed a new definition
%for the topological susceptibility $\bar{\chi_t}$, 
%which is defined in a sub-domain
%of the volume, and contains a $1/V$ correction term which is expected
%to cancel the bias of the global topological charge in the Monte Carlo history.

%Our preliminary lattice results for $\bar{\chi_t}$
%shows a clear reduction of the auto-correlation time compared 
%to that of the global topology and the $1/V$ correction term
%to cancel the global topology's bias looks working as expected. 
%Moreover, the quark mass dependence of $\bar{\chi_t}$
%shows a remarkable agreement with the prediction of 
%chiral perturbation theory.
%$\bar{\chi_t}$ is a purely gluonic quantity, nevertheless,
%it clearly reflects the quantum effect of the dynamical quarks.\\

%We thank...

\end{document}